\documentclass[runningheads]{llncs}

\usepackage[T1]{fontenc}
\usepackage{graphicx}
\usepackage{booktabs}
\usepackage[misc]{ifsym}
\usepackage{algorithm}
\usepackage{algorithmic}
\usepackage{amsmath}
\usepackage{booktabs}
\usepackage{soul}
\usepackage{float}
\usepackage{cite}
\usepackage{multirow}
\usepackage{adjustbox}
\usepackage{comment}
\usepackage{wrapfig}
\usepackage[table]{xcolor}
\usepackage{caption} 
\usepackage{subcaption}
\usepackage{enumitem}
\usepackage{hyperref}
\newcommand{\corr}{(\Letter)}
\usepackage{mwe}

\begin{document}

\newcommand{\ttl}{Density-aware Walks for Coordinated Campaign Detection}
\title{\ttl}

\titlerunning{Density-aware Walks for Coordinated Campaign Detection}

\author{Atul Anand Gopalakrishnan\inst{1} \and
Jakir Hossain\inst{1} \and
Tu\u{g}rulcan Elmas\inst{2} \and
Ahmet Erdem Sar{\i}y\"{u}ce\inst{1}\corr}

\authorrunning{A.A. Gopalakrishnan et al.}

\institute{University at Buffalo \email{\{atulanan,mh267,erdem\}@buffalo.edu}
\and
University of Edinburgh \email{telmas@ed.ac.uk}
}
\tocauthor{Atul Anand Gopalakrishnan, Jakir Hossain, Tu\u{g}rulcan Elmas, Ahmet Erdem Sar{\i}y\"{u}ce}
\toctitle{Atul Anand Gopalakrishnan, Jakir Hossain, Tu\u{g}rulcan Elmas, Ahmet Erdem Sar{\i}y\"{u}ce}
\maketitle              

\begin{abstract}
Coordinated campaigns frequently exploit social media platforms by artificially amplifying topics, making inauthentic trends appear organic, and misleading users into engagement. Distinguishing these coordinated efforts from genuine public discourse remains a significant challenge due to the sophisticated nature of such attacks.
Our work focuses on detecting coordinated campaigns by modeling the problem as a graph classification task. We leverage the recently introduced Large Engagement Networks (LEN) dataset, which contains over 300 networks capturing engagement patterns from both fake and authentic trends on Twitter prior to the 2023 Turkish elections. The graphs in LEN were constructed by collecting interactions related to campaigns that stemmed from ephemeral astroturfing.
Established graph neural networks (GNNs) struggle to accurately classify campaign graphs, highlighting the challenges posed by LEN due to the large size of its networks. To address this, we introduce a new graph classification method that leverages the density of local network structures. We propose a random weighted walk (RWW) approach in which node transitions are biased by local density measures such as degree, core number, or truss number. These RWWs are encoded using the Skip-gram model, producing density-aware structural embeddings for the nodes.
Training message-passing neural networks (MPNNs) on these density-aware embeddings yields superior results compared to the simpler node features available in the dataset, with nearly a 12\% and 5\% improvement in accuracy for binary and multiclass classification, respectively. Our findings demonstrate that incorporating density-aware structural encoding with MPNNs provides a robust framework for identifying coordinated inauthentic behavior on social media networks such as Twitter.

\end{abstract}

\section{Introduction}
Social media platforms like Twitter (now X) provide a space for people to express their opinions and stay informed about trending topics. However, like other social media platforms, Twitter is vulnerable to manipulation by malicious actors. These actors often engage in coordinated attacks that artificially amplify trends using fake accounts and bots. They can operate in a synchronized manner while concealing their identities, misleading users, journalists, and policymakers about what is genuinely trending. Such tactics also coerce users into engaging with fabricated trends, making it increasingly difficult to distinguish between organic trends and those driven by manipulation. Prior research has shown that coordinated campaigns are prevalent in several countries, including Turkey, Pakistan, and India~\cite{elmas2021ephemeral, jakesch2021trend, kausar2021towards}.

Gopalakrishnan et al.~\cite{Gopalakrishnan2025LargeEN} recently introduced a new graph classification dataset, LEN, consisting of engagement networks including some coordinated campaigns within Turkey’s Twitter sphere during the 2023 Turkish elections. To identify ground-truth campaign graphs, they focus on ephemeral astroturfing, a tactic where a coordinated network of bots rapidly generates a large volume of tweets to manipulate Twitter's trending list, only to delete them shortly afterward. In each engagement graph, nodes represent users, while edges represent user interactions in the form of retweets, quotes, or replies.

The problem of coordinated campaign detection can be considered as a graph classification task, making it well-suited for message-passing neural networks (MPNNs)~\cite{kipf2016semi, velivckovic2018graph, hamilton2017inductive, xu2018powerful}. However, Gopalakrishnan et al.'s analysis using established MPNNs highlights the challenges posed by LEN due to its large network sizes. MPNNs are often designed for domains with significantly smaller graphs, such as molecular structures. In contrast, LEN contains approximately ten times more edges, on average, than typical datasets of graphs, such as ogbn-ppa, one of the largest biological graph datasets.

\begin{figure}[!t]
    \centering
    \includegraphics[width=\linewidth]{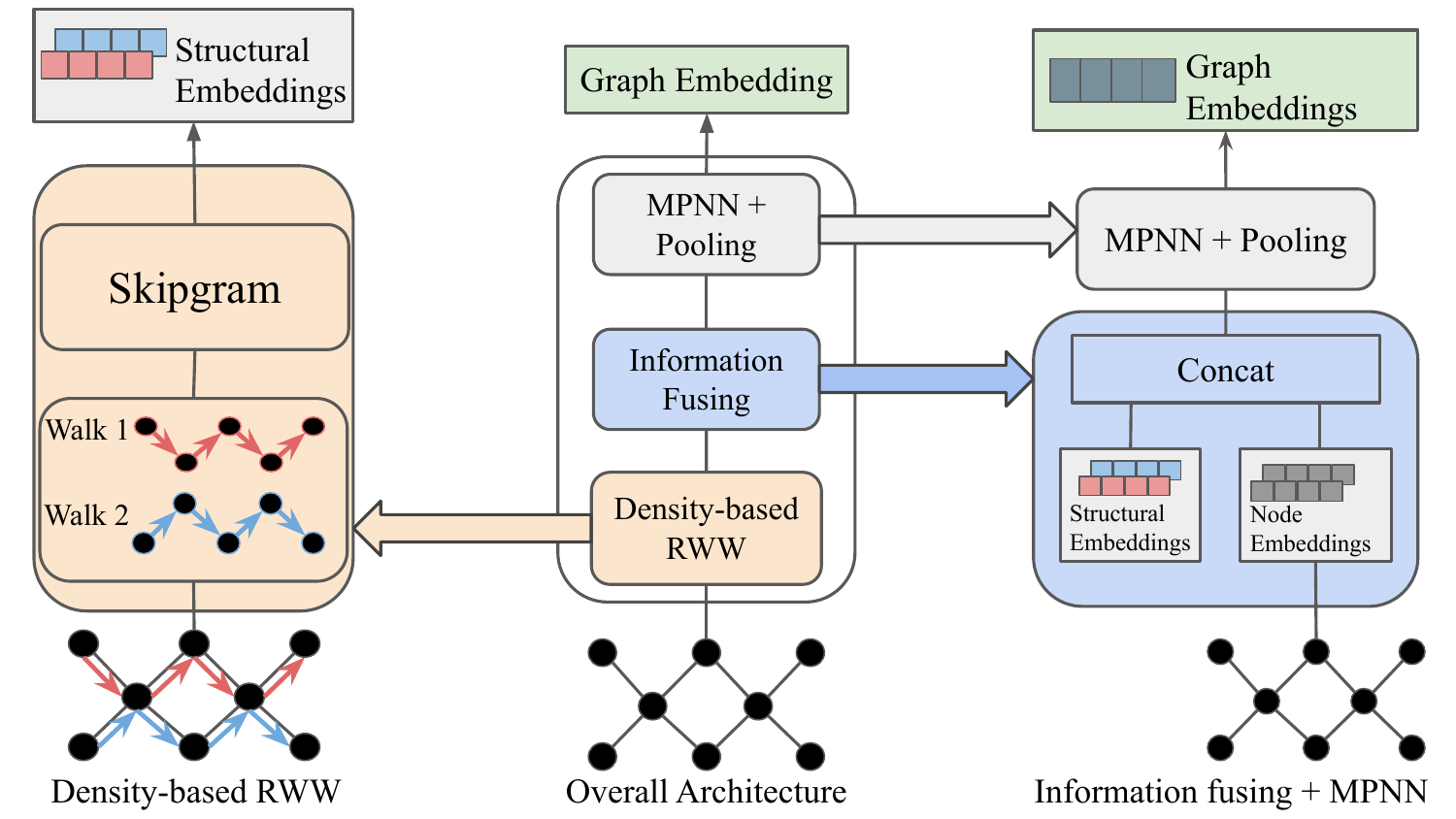}
    \caption{An overview of DECODE. Density-based random weighted walk captures the local densities around a node. This representation is then concatenated with the input node feature available in the dataset. The concatenated embedding is encoded using an MPNN and subsequently aggregated to form the graph embedding, which is used for downstream classification.}
    \label{fig:prop_fw}
\end{figure}

\textbf{Present work.} In this paper, we exploit the fact that campaign-related engagement graphs tend to be denser. We aim to accurately identify coordinated campaigns using our method, called 
\textbf{DE}nsity-aware walks for \textbf{CO}ordinated campaign \textbf{DE}tection (\textbf{DECODE}). We incorporate network density into node embeddings by leveraging node-level density properties, such as degree, core number, and truss number, using random weighted walks (RWWs). For the RWW, we sample a new node using the current node's density, ensuring that each node maintains a similar local density throughout the walk. These RWWs are converted to density-aware embeddings embeddings using Skipgram~\cite{mikolov2013efficient}. We then train a message-passing neural network (MPNN) using these embeddings as input features, enabling the model to leverage density awareness for improved classification. Figure \ref{fig:prop_fw} provides a descriptive diagram of our framework. The key contributions of our work can be summarized as follows:

\begin{itemize}
    \item We leverage multiple density measures, namely degree, core numbers and truss numbers, to distinguish campaign and non-campaign networks based on local density.
    \item We introduce DECODE, which uses RWWs to encode each node such that its embedding closely resembles those of neighboring nodes with similar densities.
    \item We train MPNNs on the LEN dataset using the the density-aware embeddings to identify campaigns and  their subtypes. To evaluate their effectiveness, we compare our models with the baselines from \cite{Gopalakrishnan2025LargeEN}.
\end{itemize}

The rest of the paper is structured as follows. In Section \ref{sec: rel_work}, we review related work, while Section \ref{sec:preliminaries} introduces the dataset and essential terminologies.  In Section \ref{sec:methodologies}, we present our methodology, including the random weighted walk algorithm and its density-awareness encoding using degrees, core numbers, and truss numbers. We verify and discuss the performance improvements using the density-aware embeddings to demonstrate their importance in Section \ref{sec:experiments}. Finally, we conclude by summarizing our findings and addressing potential limitations and future directions in Section \ref{sec:conclusion}. The code for DECODE is available at \href{https://github.com/erdemUB/ECMLPKDD25}{https://github.com/erdemUB/ECMLPKDD25}.

\section{Related Work}
\label{sec: rel_work}

In this section we discuss related work in the domain of structural and positional encoding and its relevance in MPNNs and graph transformers (Section \ref{sec:struc_pos_enc}), and coordinated campaigns on social media platforms (Section \ref{sec:trend_manipulation}).

\subsection{Structural and Positional Encoding}
\label{sec:struc_pos_enc}
Structural encoding is the process of ensuring that nodes with similar structural roles in a graph have similar embeddings. Positional encoding captures the proximity between two nodes in a graph. Both of these encodings can be obtained using random weighted walks. In Deepwalk, random walks are generated for each node~\cite{perozzi2014deepwalk}. These walks are converted to node embeddings using Skipgram~\cite{mikolov2013efficient}. Node2Vec biases the random walks to preserve both local (BFS) and global structure (DFS) ~\cite{grover2016node2vec}. Struct2Vec constructs similarity graphs using time-warping as the similarity function ~\cite{ribeiro2017struc2vec}.  Random-walks are performed on the similarity graphs to generate node embeddings. Modularized non-negative matrix factorization (M-NMF) preserves community structure by using a community embedding matrix and community modularity score in addition to random walk embeddings~\cite{wang2017community}. Random walk features have also been used to improve structural awareness of MPNNs ~\cite{zeng2023rethinking, zhou2023facilitating} and graph transformers ~\cite{rampavsek2022recipe, dwivedi2020generalization, dwivedi2022long}. Other commonly used structural and positional encoding include heat kernel~\cite{kreuzer2021rethinking, feldman2022weisfeiler}, subgraphs~\cite{bouritsas2022improving, chen2022structure}, shortest distance~\cite{li2020distance}, and node degree centralities~\cite{ying2021transformers}. In this work, we devise a new random-walk that captures the local density around nodes.

\subsection{Coordinated Campaigns on Social Media}
\label{sec:trend_manipulation}
The process by which users on a social media platform coordinate in large groups to engage in malicious behavior is known as a coordinated campaign (also known as influence operations) ~\cite{pamment2022attributing}. These coordinated campaigns are often designed to mislead users by disseminating misinformation or by propagating falsified ideologies. Examples of coordinated campaigns include, using advertisements and influencers to dominate trends~\cite{ong2018architects}, deploying bots to boost user popularity~\cite{elmas2023analyzing, elmas2022characterizing}, and state-sponsored influence operations, such as Russia's interference in the 2016 US elections~\cite{zannettou2019disinformation} and alleged coordinated attacks by the Chinese Communist Party to sway public opinion~\cite{jacobs2024whatisdemocracy}. Our work leverages coordinated campaigns driven by ephemeral astroturfing, where bots flood Twitter with random tweets to bypass filters and then delete them immediately~\cite{elmas2021ephemeral, Gopalakrishnan2025LargeEN}. Since Twitter updates trends in windows, deleted tweets remain unaccounted for until the next update, allowing adversaries to exploit the illusion of organic engagement.

Over the years, methods have been developed to counter coordinated efforts on Twitter. These include techniques such as tweet and hashtag similarity~\cite{luceri2024unmasking, pacheco2020unveiling}, temporal methods focusing on tweet frequency \cite{pacheco2020unveiling, vishnuprasad2024tracking}, shared URLs and articles \cite{gabriel2023inductive}, and detecting other coordination signals~\cite{erhardt2024hidden, weber2021amplifying,pote2024coordinated}. Recent approaches have explored centrality-based node pruning on similarity networks~\cite{luceri2024unmasking}, and graph neural networks~\cite{elmas2024teamfollowback,gabriel2023inductive} for detecting these attacks. Our work attempts to identify coordinated campaigns by modeling it as a graph classification problem. Additionally, we encode density-based properties using RWWs. Incorporating density-aware embeddings into MPNN training leads to improved performance in classification compared to using only the raw node features from the dataset.

\section{Preliminaries}
\label{sec:preliminaries}

In this section, we describe our dataset and the key terminologies required for this work. We first provide details on the LEN dataset in Section \ref{sec:astroturf}. Then we give a brief overview of the notation, the density metrics used (degree, $k$-core, $k$-truss) and message passing neural networks in Section \ref{sec:graphs}.

\subsection{Large Engagement Networks (LEN) Dataset}
\label{sec:astroturf}
Large Engagement Networks (LEN) is graph dataset that contains coordinated campaigns related to Turkish Twitter. It focuses on the 2023 elections in Turkey when this issue was prevalent. The campaign graphs in the dataset are an outcome of ephemeral astroturfing. The dataset comprises of 314 engagement networks, where each network is associated to a trend. There are 179 campaign graphs and 135 non-campaign graphs. These graphs are further divided into sub-types such as politics, news, finance and more, as shown in Table \ref{tab:dataset_properties}. The nodes represent users and edges represents engagements between the users. A directed edge from node $X$ to $Y$, signifies that $X$ engaged with (retweeted, replied to, or quoted) $Y$. The graphs also consist of node and edge features. The features used for the node attributes include user description (bio), follower count, following count, user's total tweet count, and user's verification status. The edge attributes include the type of engagement (retweet, reply, or quote), engagement count (e.g., number of retweets), impression count, text,  number of likes, whether the tweet is labeled as sensitive or not, and the timestamp of the tweet. Gopalakrishnan et al. provides three benchmarks for the LEN dataset: (1) binary classification to classify the networks into campaign and non-campaign networks; (2) multi-class classification to categorize campaigns into one of the 7 sub-types as shown in Table \ref{tab:dataset_properties}; and (3) binary classification of news networks into campaign and non-campaign. We use the LEN dataset as it is the only ground-truth graph classification dataset that identifies if a trend's popularity is driven by coordinated campaigns.

\begin{table}[!t]
    \centering
    \caption{Statistics of the engagement networks for LEN, containing 314 networks.}
    \vspace{2ex}
    \fontsize{9pt}{11pt}\selectfont
    \begin{adjustbox}{max width=\textwidth}
    \setlength{\tabcolsep}{1.5pt}
    \begin{tabular}{l|l|r|r|r|r|r|r|r|l}
    & \fontsize{10}{12}\selectfont \textbf{Sub-types} 
    & \fontsize{10}{12}\selectfont \textbf{\# G} 
    & \multicolumn{3}{|c|}{\fontsize{10}{12}\selectfont \textbf{\# nodes}} 
    & \multicolumn{3}{|c|}{\fontsize{10}{12}\selectfont \textbf{\# edges}} 
    & \fontsize{10}{12}\selectfont \textbf{Explanation}\\

    \hline
         & & & \fontsize{10}{12}\selectfont \textbf{Max} & \fontsize{10}{12}\selectfont \textbf{Min} & \fontsize{10}{12}\selectfont \textbf{Avg} & \fontsize{10}{12}\selectfont \textbf{Max} & \fontsize{10}{12}\selectfont \textbf{Min} & \fontsize{10}{12}\selectfont \textbf{Avg} & \\
    \hline
        \multirow{8}{*}{\rotatebox[origin=c]{90}{Campaign}} 
        & Politics & 62 & 50,286 & 100 & 6,570 & 71,704 & 203 & 10,210 & {\fontsize{9}{11}\selectfont Political content including slogans, and misinformation camps.}\\ 
        & Reform & 58 & 19,578 & 131 & 1,229 & 1,105,918 & 540 & 25,268 & {\fontsize{9}{11}\selectfont Organized movements advocating political changes.}\\ 
        & News & 24 & 54,996 & 581 & 10,368 & 80,784 & 942 & 15,582 & {\fontsize{9}{11}\selectfont News amplified through bot and troll activity.}\\ 
        & Finance & 14 & 9,976 & 273 & 1,802 & 10,725 & 243 & 2,334 & {\fontsize{9}{11}\selectfont Financial promotions, primarily cryptocurrency-related.}\\ 
        & Noise & 9 & 55,933 & 454 & 12,180 & 48,937 & 473 & 10,882 & {\fontsize{9}{11}\selectfont Content that does not fit into any specific category.}\\ 
        & Cult & 6 & 7,880 & 313 & 2,303 & 11,615 & 637 & 3,431 & {\fontsize{9}{11}\selectfont Slogans from a cult leveraging bots.}\\ 
        & Entertainment & 3 & 4,220 & 678 & 2,237 & 132,013 & 3,806 & 48,767 & {\fontsize{9}{11}\selectfont Celebrities using bots for self-promotion.}\\ 
        & Common & 3 & 9,974 & 3,487 & 5,919 & 9,470 & 2,818 & 7,066 & {\fontsize{9}{11}\selectfont Frequently used phrases forming trends organically}\\ 
        & {\bf Overall} & {\bf 179} & {\bf 55,933} & {\bf 100} & {\bf 5,157} & {\bf 1,105,918} & {\bf 203} & {\bf 16,006} &  \\
    \hline
        \multirow{9}{*}{\rotatebox[origin=c]{90}{Non-Campaign}} 
        & News & 52 & 95,575 & 818 & 24,834 & 213,444 & 709 & 43,201 & {\fontsize{9}{11}\selectfont Coverage of news not sourced from Twitter.}\\ 
        & Sports & 30 & 75,653 & 469 & 9,530 & 101,656 & 403 & 12,948 & {\fontsize{9}{11}\selectfont Discussions around major sports events.}\\ 
        & Festival & 17 & 119,952 & 885 & 35,466 & 199,305 & 803 & 55,947 & {\fontsize{9}{11}\selectfont Trends related to holidays, festivals, and special occasions.}\\ 
        & Internal & 11 & 87,720 & 4,188 & 33,061 & 196,103 & 4,374 & 54,442 & {\fontsize{9}{11}\selectfont Events primarily generated within Twitter's ecosystem.}\\ 
        & Common & 10 & 64,320 & 1,214 & 17,079 & 99,306 & 1,270 & 24,869 & {\fontsize{9}{11}\selectfont Frequently used phrases forming trends organically.}\\ 
        & Entertainment & 8 & 20,060 & 1,477 & 7,289 & 45,211 & 1,712 & 12,578 & {\fontsize{9}{11}\selectfont Engagement with popular TV shows and online videos.}\\ 
        & Announ. cam. & 4 & 26,358 & 6,650 & 13,382 & 50,864 & 14,362 & 24,817 & {\fontsize{9}{11}\selectfont Officially launched political campaigns.}\\ 
        & Sports cam. & 3 & 4,661 & 2,880 & 3,654 & 7,367 & 4,451 & 5,534 & {\fontsize{9}{11}\selectfont Hashtags initiated by professional sports teams.}\\ 
        & {\bf Overall} & {\bf 135} & {\bf 119,952} & {\bf 469} & {\bf 20,632} & {\bf 213,444} & {\bf 403} & {\bf 33,765} &  \\
    \end{tabular}
    \end{adjustbox}
    \label{tab:dataset_properties}    
\end{table}

\subsection{Notation, Density Metrics, and MPNNs}
\label{sec:graphs}
A graph is a collection of vertices and edges. It is depicted as $G = (V, E)$, where $V$ is the number of vertices and $E$ is the number of edges. A graph can also be represented as $G = (A, X, y)$, where $A$ is the adjacency matrix, $X$ is the feature matrix for the nodes, and $y$ is the graph's label.

\subsubsection{Degree, $k$-core and $k$-truss:} The degree of a node is the number of edges connected to it, providing a simple measure of its local connectivity~\cite{freeman1978centrality}. A $k$-core is a subgraph in which every node has at least $k$ connections within the subgraph \cite{seidman1983network}. The core number of a node represents the largest $k$-core to which it belongs. Computing core numbers of all the nodes in a graph has a linear cost, $O(|E|)$. Similarly, a $k$-truss is defined as a subgraph where each edge is part of at least $k-2$ triangles within the subgraph \cite{cohen2008trusses}. The truss number of an edge indicates the highest $k$-truss to which it belongs. Computing truss numbers is a bit costly, $O(|E|^{1.5})$, but is still polynomial and practical for large networks. Since truss numbers are edge-based, we compute a node’s truss number by averaging the truss numbers associated to the connected edges. Degree, core number, and truss number all measure graph density, with degree indicating direct connections, and core and truss numbers reflecting a node’s role in dense substructures. 

We use three local density measures for a node: (1) degree of the node, (2) core number of the node, and (3) average truss numbers of all edges incident to the node (we simply refer them as degree, core, and truss number of a node in the rest of the paper). Note that we ignore the edge directions in the engagement networks, hence we use the original definitions of $k$-core and $k$-truss for undirected graphs.

\subsubsection{Message Passing Neural Networks:} MPNNs consist of two steps, aggregate and update, as shown in Equation \ref{eq: modified_msg}, where $\mathcal{N}(v)$ is used to represent the neighborhood of node $v$.

\begin{equation}
\label{eq: modified_msg}
h_v^{(l+1)} = \text{UPDATE} \left( h_v^{(l)} , \, \text{AGGREGATE} \left(\{ h_u^{(l)} \mid u \in \mathcal{N}(v) \} \right) \right)
\end{equation}

In the aggregate step, each node gathers information from its neighbors. This typically involves summing, averaging, or applying more complex functions (e.g., attention mechanisms) to the neighbors’ feature vectors. In the update step, the aggregated information is combined with the node’s own features to update its representation. To do so, a neural network (example, an MLP) or a simple transformation (example, a weighted sum) is applied. Therefore, the nodes refine their representations based on the information received. MPNNs generally differ in the aggregation strategy used. GCN uses dual-degree normalization to account for the varying number of neighbors each node may have~\cite{kipf2016semi}. GAT uses attention weight to assign varying weights to each neighbor~\cite{velivckovic2018graph}. GIN uses an MLP to perform aggregation using a trainable parameter ($\epsilon$) to determine the amount of importance given to the ego node, as compared to its neighbors~\cite{xu2018powerful}. GraphSAGE is an inductive graph representational learning model that has the  ability to generalize to unseen nodes, unlike transductive models~\cite{hamilton2017inductive}. This is done by learning a message-passing model on a sampled set of nodes in the given graph.

\section{Methodology}
\label{sec:methodologies}

\begin{table}[!t]
    \centering
    \caption{Descriptive statistics to indicate the average degree, core number and truss numbers of the nodes across different sub-types spanning campaign and non-campaign graphs. Overall, campaign graphs exhibit higher local densities than the non-campaign ones.}
    \vspace{2ex}
    \begin{adjustbox}{width=\textwidth} 
        \begin{tabular}{l|l|c|c|c}
        & \textbf{Sub-types} & \textbf{Mean Degree} & \textbf{Mean Core number} & \textbf{Mean Truss number} \\
        \hline
        \multirow{8}{*}{\rotatebox[origin=c]{90}{Campaign}} 
         & Politics & 3.108 $\pm$ 1.483 & 1.635 $\pm$ 0.872 & 0.384 $\pm$ 1.595 \\
         & Reform & 16.258 $\pm$ 12.118 & 9.067 $\pm$ 7.067 & 17.104 $\pm$ 10.545 \\
         & News & 3.015 $\pm$ 0.990 & 1.598 $\pm$ 0.481 & 0.172 $\pm$ 0.365 \\
         & Finance & 2.590 $\pm$ 1.512 & 1.414 $\pm$ 0.823 & 0.287 $\pm$ 0.977 \\
         & Noise & 2.080 $\pm$ 0.790 & 1.214 $\pm$ 0.342 & 0.261 $\pm$ 0.370 \\
         & Cult & 2.981 $\pm$ 0.690 & 1.580 $\pm$ 0.409 & 0.569 $\pm$ 0.919 \\
         & Entertainment & 11.477 $\pm$ 0.264 & 6.180 $\pm$ 0.171 & 2.311 $\pm$ 1.920 \\
         & Common & 2.391 $\pm$ 1.586 & 1.380 $\pm$ 0.793 & 0.572 $\pm$ 0.751 \\
         & \textbf{Overall} & \textbf{3.941 $\pm$ 10.459} & \textbf{2.120 $\pm$ 6.083} & \textbf{5.057 $\pm$ 7.933} \\
        \hline
        \multirow{9}{*}{\rotatebox[origin=c]{90}{Non-Campaign}} 
         & News & 3.443 $\pm$ 0.845 & 1.778 $\pm$ 0.412 & 0.179 $\pm$ 0.106 \\
         & Sports & 2.718 $\pm$ 0.581 & 1.428 $\pm$ 0.242 & 0.155 $\pm$ 0.150 \\
         & Festival & 2.867 $\pm$ 0.501 & 1.526 $\pm$ 0.239 & 0.245 $\pm$ 0.157 \\
         & Internal & 3.293 $\pm$ 0.901 & 1.705 $\pm$ 0.438 & 0.184 $\pm$ 0.110 \\
         & Common & 2.913 $\pm$ 0.650 & 1.529 $\pm$ 0.305 & 0.109 $\pm$ 0.068 \\
         & Entertainment & 3.453 $\pm$ 0.941 & 1.816 $\pm$ 0.471 & 0.651 $\pm$ 0.491 \\
         & Announced Camp. & 3.711 $\pm$ 1.135 & 1.925 $\pm$ 0.588 & 1.309 $\pm$ 0.970 \\
         & Sports Camp. & 3.029 $\pm$ 0.191 & 1.590 $\pm$ 0.100 & 0.037 $\pm$ 0.012 \\
         & \textbf{Overall} & \textbf{3.210 $\pm$ 0.81} & \textbf{1.672 $\pm$ 0.387} & \textbf{0.219 $\pm$ 0.261}\\
        \hline
        \end{tabular}
    \end{adjustbox}
    \label{tab:density_comparison}
\end{table}

We propose DECODE, a random weighted walk (RWW) approach for learning density-aware node embeddings. Here, node densities are used to determine transition probabilities in the RWWs. This emphasis on density is because campaign graphs in LEN are denser than non-campaign graphs. Specifically, we use degree, core number, and truss number as density metrics due to their widespread use and computational efficiency~\cite{batagelj2003m, sariyuce2017nucleus}. Table \ref{tab:density_comparison} provides detailed statistics showcasing the density metrics across campaign and non-campaign graphs. Notably, the densest campaign graphs belonged to the reform sub-type, which constitutes a large portion of the dataset, as shown in Table \ref{tab:dataset_properties}.

\begin{algorithm}[!t]
\caption{Density-aware random weighted walk (DECODE)}
\label{alg: rww}
\begin{algorithmic}
\STATE \textbf{Input:} Graph \( G = (V, E) \), walk length \( L \), density func. \( \phi: V \to [0, 1] \), threshold \(\tau\)
\STATE \textbf{Output:} List of walks \( W \)
\
\STATE Initialize \( W \gets [] \)
\FOR{each node \( v \in V \)}
    \STATE Initialize walk \( w \gets [v] \)
    \FOR{\( t = 1 \) to \( L \)}
        \STATE \(v_t\) = Top (\(w\))
        \STATE Let \( N(v_t) \gets \{ u \in V \mid (v_t, u) \in E \} \)
        \IF{\( N(v_t) \neq \emptyset \)}
            \IF{\( \phi(v_t) > \tau \)}
                \STATE Set \( w_u = \phi(u) \) for all \( u \in N(v_t) \)
            \ELSE
                \STATE Set \( w_u = {1 - \phi(u)} \) for all \( u \in N(v_t) \)
            \ENDIF
            \STATE Sample \( v_{t+1} \sim P(u) = {w_u}/{\sum_{u' \in N(v_t)} w_{u'}} \)
            \STATE Push \( v_{t+1} \) to \( w \)
        \ENDIF
    \ENDFOR
    \STATE Append \( w \) to \( W \)
\ENDFOR
\RETURN \( W \)
\end{algorithmic}
\end{algorithm}

Algorithm \ref{alg: rww} provides a formal overview of DECODE. In our algorithm, $\phi$ represents the density function, where $\phi(v)$ returns the normalized density of a given node $v$. The function $\phi$ is defined based on the chosen density metric for RWWs. It can be set to return the degree, core number, or truss number of a node.

Additionally, we introduce $\tau$, a scalar threshold parameter that differentiates between high and low-density nodes in RWWs. The threshold is set to one of the following values: 0.5, the median node density in the graph, or the midpoint of node densities, as detailed in Section \ref{sec:exp_setup}. The steps for collecting RWWs in our algorithm are as follows:

\begin{enumerate}
    \item At each step of the RWW, the next node is selected based on the density of the current node. 
    \item If the current node's density exceeds the threshold $\tau$, transitions to higher-density neighbors are preferred, with sampling weights defined as (\( w_u = \phi(u) \)), where $w_u$ represents the weight assigned to node $u$
    \item Conversely, if the current node's density is below $\tau$, transitions to lower-density neighbors are favored by inverting the sampling weights (\( w_u = 1-\phi(u) \)).
    \item The transition probabilities for the neighbors are obtained by normalizing the sampling weights and new nodes are sampled using them at each step.
    \item Once we obtain the RWWs, we use Skipgram to encode them, following prior methods~\cite{perozzi2014deepwalk, grover2016node2vec}.
\end{enumerate}

These density-aware embeddings and node feature are fed into the MPNNs for downstream classification. The MPNNs used in this paper include GCN, GAT, GIN, and GraphSAGE. In the following section we discuss the experimental setup used in this paper and discuss our results for binary and multiclass classification by comparing our method to the results provided in~\cite{Gopalakrishnan2025LargeEN}.

\section{Experimental Evaluation}
\label{sec:experiments}

We evaluate the performance of DECODE on the LEN dataset using two tasks: (i) campaign vs. non-campaign classification in engagement networks (binary classification) and (ii) campaign sub-type classification, where the sub-types are provided in Table \ref{tab:dataset_properties} (multi-class classification). Section \ref{sec:exp_setup} details the experimental setup. Sections \ref{sec:bin_class} and \ref{sec:mult_class} present the experimental results for binary and multi-class classification, respectively.

\subsection{Experimental Setup}
\label{sec:exp_setup}

We run our model on two input configurations: (i) density-aware embeddings and (ii) a concatenation of density-aware embeddings with the input node features available in the dataset. We consider each of the three density-based features in our random walks—degrees, core numbers, and truss numbers—and provide comparisons. To contextualize the empirical results of DECODE, we compare our method against four baselines: GCN, GAT, GIN, and GraphSAGE. These models are trained solely on the input node features available in the dataset. This comparison allows us to evaluate the importance of density-aware embeddings over existing node features. To construct the RWW embeddings, we set the walk length to 100. For encoding the nodes using Skipgram, we use a window length of 4, meaning each node is encoded using its four neighboring nodes in the random weighted walks. The walk embedding size is set to 128. We set the threshold parameter ($\tau$) to the following values:

\begin{itemize}
    \item \textbf{0.5}: A fixed value of 0.5.
    \item \textbf{Median}: The median of the list of the density-based features in a graph.
    \item \textbf{Mid-point (abbreviated as mid)}: This value is calculated as the average of the smallest and largest values of the density-based feature under consideration.
\end{itemize}

For MPNNs, we perform hyperparameter tuning over hidden layer sizes, \( h \in \{128, 256, 512, 1024\} \), and learning rates, \( l \in \{0.001, 0.0001, 0.00001\} \) as done in \cite{Gopalakrishnan2025LargeEN}. We also use mean pooling to produce graph embeddings.

\begin{table}[!t]
    \centering
    \caption{Accuracy for binary classification. \textbf{NF} denotes the MPNN trained with node features and \textbf{RWW} denotes the one that used random-weighted walks. Best in each group is in bold. Underlined value denotes the best overall accuracy.}
        \vspace{2ex}
    \begin{adjustbox}{max width=\textwidth}
    \begin{tabular}{c | c | c | c | c | c | c| c}
        
        \multirow{2}{*} & \multirow{2}{*}{\textbf{Input}} & \multicolumn{2}{c|}{\textbf{Degree}} & 
        \multicolumn{2}{c|}{\textbf{Core Numbers}} & 
        \multicolumn{2}{c}{\textbf{Truss Numbers}} \\
        & & \textbf{$\tau$} & \textbf{Accuracy}& \textbf{$\tau$} & \textbf{Accuracy}& \textbf{$\tau$} & \textbf{Accuracy}\\ 

        \hline
        
        \multirow{3}{*}{GCN} & NF & - & 0.702 ± 0.018 & - & 0.702 ± 0.018 & - & 0.702 ± 0.018 \\
        & RWW & mid & \textbf{0.810 ± 0.013} & 0.5 & 0.787 ± 0.010 & median & 0.800 ± 0.019 \\
        & NF + RWW & mid & 0.784 ± 0.006 & median & \textbf{0.805 ± 0.018} & 0.5 & \textbf{0.803 ± 0.019} \\
        
        \hline
        
        \multirow{3}{*}{GAT} & NF & - & 0.735 ± 0.015 & - & 0.735 ± 0.015 & - & 0.735 ± 0.015 \\
        & RWW & median & \textbf{0.795 ± 0.015} & mid & \textbf{0.808 ± 0.022} & median & \underline{\textbf{0.836 ± 0.016}}\\
        & NF + RWW & median & 0.792 ± 0.022 & 0.5 & 0.785 ± 0.010 & median & 0.792 ± 0.030 \\
        
        \hline
        
        \multirow{3}{*}{GIN} & NF & - & 0.633 ± 0.065 & - & 0.633 ± 0.065 & - & 0.633 ± 0.065 \\
        & RWW & 0.5 & \textbf{0.756 ± 0.005} & median & 0.766 ± 0.014 & 0.5 & 0.771 ± 0.006 \\
        & NF + RWW & 0.5 &  0.751 ± 0.005 & mid & \textbf{0.792 ± 0.014} & mid & \textbf{0.782 ± 0.019}\\
        
        \hline
        
        \multirow{3}{*}{SAGE} & NF & - & 0.729 ± 0.006 & - & 0.729 ± 0.006 & - & 0.729 ± 0.006 \\
        & RWW & 0.5 & \textbf{0.852 ± 0.010} & mid & 0.774 ± 0.025 & 0.5 & \textbf{0.834 ± 0.052}\\
        & NF + RWW & median & 0.758 ± 0.010 & 0.5 & \textbf{0.790 ± 0.017} & 0.5 & 0.813 ± 0.018 \\
        
    \end{tabular}
    \end{adjustbox}
    \label{tab:binary_classification_acc}
\end{table}

\begin{table}[!t]
    \centering
    \caption{F1-score for binary classification. \textbf{NF} denotes the MPNN trained with node features and \textbf{RWW} denotes the one that used random-weighted walks. Best in each group is in bold. Underlined value denotes the best overall F1-score.}
        \vspace{2ex}
    \begin{adjustbox}{max width=\textwidth}
    \begin{tabular}{c | c | c | c | c | c | c | c}
        \multirow{2}{*} & \multirow{2}{*}{\textbf{Inp.}} & \multicolumn{2}{c|}{\textbf{Degree}} & 
        \multicolumn{2}{c|}{\textbf{Core Numbers}} & 
        \multicolumn{2}{c}{\textbf{Truss Numbers}} \\
        & & \textbf{$\tau$} & \textbf{F1} & \textbf{$\tau$} & \textbf{F1} & \textbf{$\tau$} & \textbf{F1} \\
        
        \hline
        
        \multirow{3}{*}{GCN} & NF & - & 0.687 ± 0.021 & - & 0.687 ± 0.021 & - & 0.687 ± 0.021 \\
        & RWW & mid & \textbf{0.839 ± 0.004} & median & 0.814 ± 0.008 & median & 0.806 ± 0.021 \\
        & NF + RWW & median & 0.805 ± 0.011 & median & \textbf{0.838 ± 0.017} & 0.5 & \textbf{0.824 ± 0.017} \\
        
        \hline
        
        \multirow{3}{*}{GAT} & NF & - & 0.765 ± 0.018 & - & 0.765 ± 0.018 & - & 0.765 ± 0.018 \\
        & RWW & median & \textbf{0.825 ± 0.012} & mid & \textbf{0.840 ± 0.015} & median & \textbf{0.853 ± 0.011} \\
        & NF + RWW & median & 0.820 ± 0.020 & 0.5 & 0.824 ± 0.006 & median & 0.824 ± 0.030 \\
        
        \hline
        
        \multirow{3}{*}{GIN} & NF & - & 0.710 ± 0.037 & - & 0.710 ± 0.037 & - & 0.710 ± 0.037 \\
        & RWW & 0.5 & \textbf{0.807 ± 0.005} & median & 0.800 ± 0.013 & 0.5 & 0.790 ± 0.003 \\
        & NF + RWW & mid & 0.782 ± 0.006 & mid & \textbf{0.816 ± 0.012} & mid & \textbf{0.795 ± 0.019} \\
        
        \hline
        
        \multirow{3}{*}{SAGE} & NF & - & 0.713 ± 0.008 & - & 0.713 ± 0.008 & - & 0.713 ± 0.008 \\
        & RWW & 0.5 & \underline{\textbf{0.877 ± 0.010}} & mid & 0.789 ± 0.020 & 0.5 & \textbf{0.857 ± 0.031} \\
        & NF + RWW & median & 0.803 ± 0.007 & 0.5 & \textbf{0.820 ± 0.016} & 0.5 & 0.834 ± 0.011 \\
    \end{tabular}
    \end{adjustbox}
    \label{tab:binary_classification_f1}
\end{table}

\subsection{Results for Campaign vs Non-campaign Classification}
\label{sec:bin_class}
\begin{figure}[!t]
    \centering
    \includegraphics[width=\textwidth]{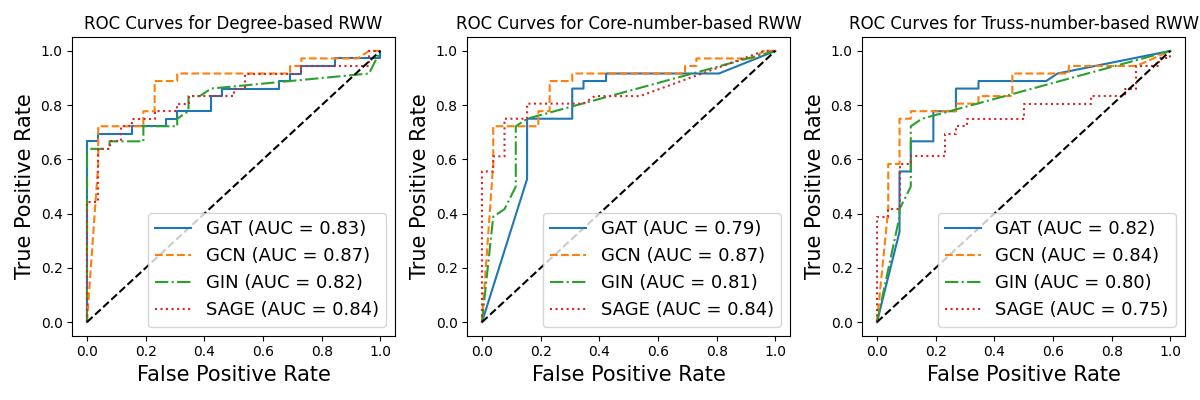}
    \caption{Receiver Operating Characteristic (ROC) curves for campaign vs. non-campaign classification for degree, core-number and truss-number based random weighted walks. The best performing input format and threshold are taken into consideration for each model.}
    \label{fig:roc_curves}
\end{figure}

LEN consists of 179 campaign graphs and 135 non-campaign graphs. The results for accuracy and F1-score are presented in Tables \ref{tab:binary_classification_acc} and \ref{tab:binary_classification_f1}, respectively. We observe the following key insights:

\begin{itemize}
    \item Pairing GraphSAGE with degree-based RWW achieves the best performance, yielding an accuracy of 0.852 $\pm$ 0.010 and an F1-score of 0.877 $\pm$ 0.010, surpassing the best baseline in \cite{Gopalakrishnan2025LargeEN}  by 0.117 and 0.112 for accuracy and F1-score, respectively. The value of $\tau$ is set to 0.5 in this case.
    \item RWW features consistently outperform LEN node features, achieving higher accuracy and F1-score in most cases.
    \item Embeddings learnt from degree-based RWW generally outperforms other density-aware variants, achieving the highest AUROC scores across all models. The only exception is when GCN and GraphSAGE are trained on embeddings obtained from $k$-core-based RWW, where the AUC scores are identical. The results are illustrated in Figure \ref{fig:roc_curves}. 
    \item The best-performing threshold varies depending on the model used. Median serves as the best threshold for GCN and GAT, while 0.5 is optimal for GIN and GraphSAGE.  
\end{itemize}
The above insights suggest that RWW based methods yield improvements in performance for both accuracy and F1-score. Additionally, degree-based RWW generally outperforms core or truss-based RWWs. However the choice of threshold is model-dependent.

\subsection{Results of Campaign-type Classification}
\label{sec:mult_class}
The goal here is to classify campaign graphs into one of the seven sub-types described in Table \ref{tab:dataset_properties}. Among these, the most common categories are Politics (62 graphs) and Reform (58 graphs). The results for accuracy and macro F1-scores are provided in Tables \ref{tab:multiclass_accuracy} and \ref{tab:multiclass_macro_f1}, respectively.

\begin{figure}[t]
    \centering
    \includegraphics[width=\textwidth]{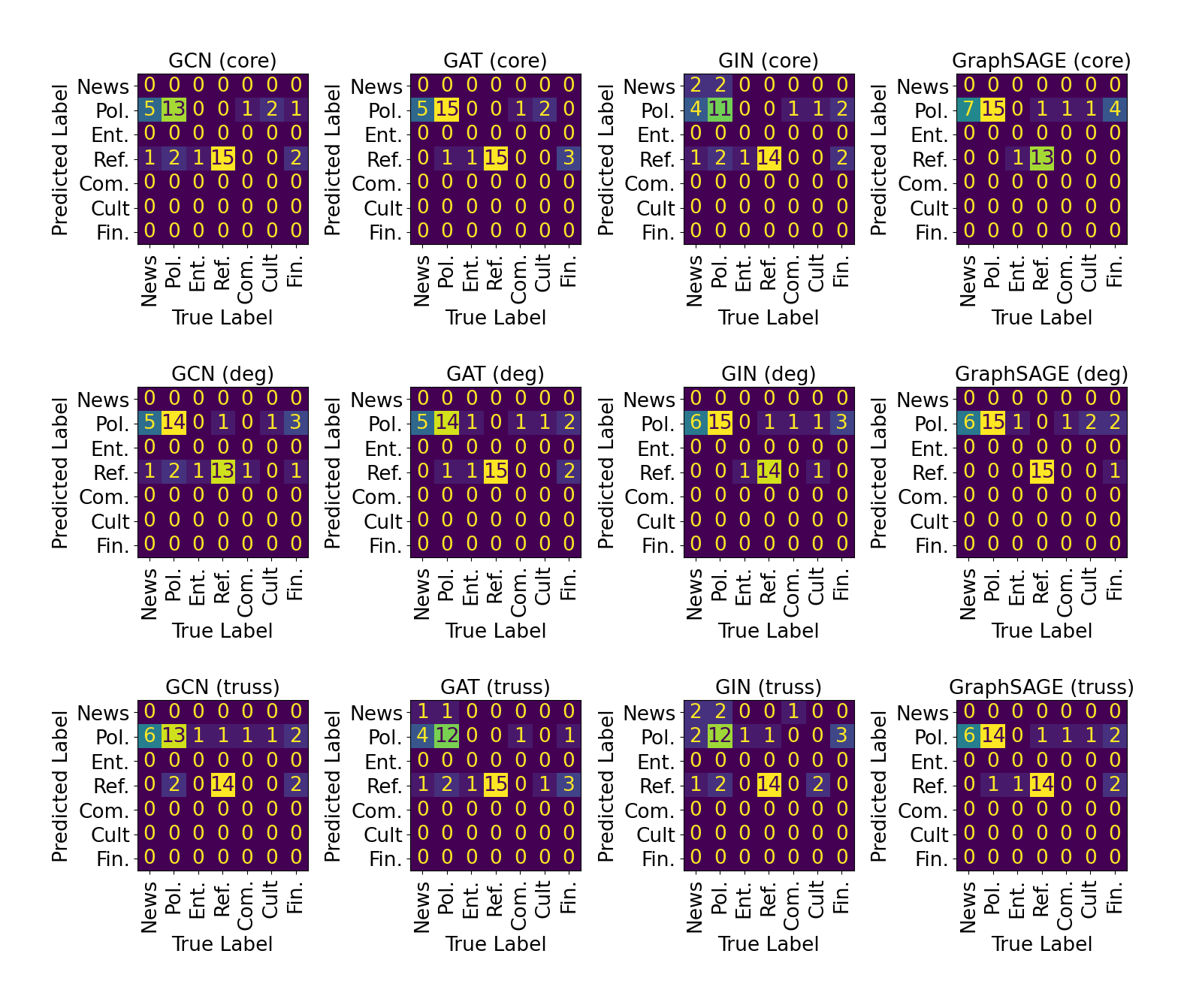}
    \caption{Confusion matrices for multi-class classification. The best performing configuration is considered for each model and density pair is displayed here.}
    \label{fig:conf_mat}
\end{figure}

From these results, the following inferences can be made:
\begin{itemize}
    \item Pairing GIN with degree-based RWW achieves the best performance, with an accuracy of 0.679 $\pm$ 0.001, surpassing the baseline in \cite{Gopalakrishnan2025LargeEN} by 0.045. 
    \item The model accuracies benefit the most when input node features from the dataset are combined with density-aware embeddings, outperforming all the other setups in a majority of the scenarios.
    \item The best-performing thresholds are mid for GCN, 0.5 for GAT, and median for GIN and GraphSAGE, yielding the highest accuracy for each model.
    \item The highest macro-F1 score obtained by our work is 0.338 $\pm$ 0.051 (for GIN with truss-based RWW and $\tau$ set to 0.5) which is 0.013 less than the best performing baseline provided in \cite{Gopalakrishnan2025LargeEN}. We believe this happens due to label imbalance. Several campaign-type labels (example, finance, entertainment, cult) have very few samples, making them harder to classify.
    \item We also provide confusion matrices for the models across various RWW methods in Figure \ref{fig:conf_mat}, where we display the confusion matrix for the best-performing configuration of each model-RWW pair. From this, we again observe that models struggle to accurately classify labels with fewer graphs.
\end{itemize}

\begin{table}[!t]
    \centering
        \caption{Accuracy results for multiclass classification. \textbf{NF} denotes the MPNN trained with node features and \textbf{RWW} denotes the one that used random-weighted walks. Best in each group is in bold. Underlined value denotes the best overall accuracy.}
        \vspace{2ex}
    \begin{adjustbox}{max width=\textwidth}
    \begin{tabular}{c | c | c | c | c | c | c| c}
        \multirow{2}{*} & \multirow{2}{*}{\textbf{Inp.}} & \multicolumn{2}{c|}{\textbf{Degree}} & 
        \multicolumn{2}{c|}{\textbf{Core Numbers}} & 
        \multicolumn{2}{c}{\textbf{Truss Numbers}} \\
        & & \textbf{$\tau$} & \textbf{Acc.} & \textbf{$\tau$} & \textbf{Acc.} & \textbf{$\tau$} & \textbf{Acc.} \\ 
    
        \hline
        
        \multirow{3}{*}{GCN}
        & NF & - & 0.533 ± 0.041 & - & 0.533 ± 0.041 & - & 0.533 ± 0.041 \\
        & RWW & median & 0.619 ± 0.019 & 0.5 & 0.614 ± 0.011 & 0.5 & 0.628 ± 0.000 \\
        & NF + RWW & 0.5 & \textbf{0.647 ± 0.009} & mid & \textbf{0.665 ± 0.011} & mid & \textbf{0.651 ± 0.015} \\
        
        \hline
        
        \multirow{3}{*}{GAT}
        & NF & - & 0.567 ± 0.033 & - & 0.567 ± 0.033 & - & 0.567 ± 0.033 \\
        & RWW & 0.5 & \textbf{0.647 ± 0.009} & mid & 0.670 ± 0.037 & mid & 0.623 ± 0.017 \\
        & NF + RWW & 0.5 & 0.628 ± 0.025 & 0.5 & \textbf{0.674 ± 0.001} & mid & \textbf{0.633 ± 0.027}\\
        
        \hline
        
        \multirow{3}{*}{GIN}
        & NF & - & 0.633 ± 0.067 & - & 0.633 ± 0.067 & - & 0.633 ± 0.067 \\
        & RWW & mid & \underline{\textbf{0.679 ± 0.001}} & median & 0.647 ± 0.009 & mean & 0.637 ± 0.024 \\
        & NF + RWW & median & 0.670 ± 0.009 & median & \textbf{0.656 ± 0.017} & median & \textbf{0.679 ± 0.023} \\
        
        \hline
        
        \multirow{3}{*}{SAGE}
        & NF & - & 0.583 ± 0.053 & - & 0.583 ± 0.053 & - & 0.583 ± 0.053 \\
        & RWW & 0.5 & 0.637 ± 0.01 & 0.5 & 0.656 ± 0.027 & median & \textbf{0.660 ± 0.011} \\
        & NF + RWW & 0.5 & \textbf{0.665 ± 0.011} & median & \textbf{0.665 ± 0.065} & median & 0.651 ± 0.001 \\
    \end{tabular}
    \end{adjustbox}

    \label{tab:multiclass_accuracy}
\end{table}

\begin{table}[!t]
    \centering
        \caption{Macro F1-score results for multiclass classification. \textbf{NF} denotes the MPNN trained with node features and \textbf{RWW} denotes the one that used random-weighted walks. Best in each group is in bold. Underlined values denote the best overall Macro-F1 score.}
        \vspace{2ex}
    \begin{adjustbox}{max width=\textwidth}
    \begin{tabular}{c | c | c | c | c | c | c | c}
        
        \multirow{2}{*} & \multirow{2}{*}{\textbf{Inp.}} & \multicolumn{2}{c|}{\textbf{Degree}} & 
        \multicolumn{2}{c|}{\textbf{Core Numbers}} & 
        \multicolumn{2}{c}{\textbf{Truss Numbers}} \\
        & & \textbf{$\tau$} & \textbf{Macro-F1} & \textbf{$\tau$} & \textbf{Macro-F1} & \textbf{$\tau$} & \textbf{Macro-F1} \\ 
        
        \hline
        
        \multirow{2}{*}{GCN}
        & NF & - & \textbf{0.251 ± 0.022} & - & 0.251 ± 0.022 & - & 0.251 ± 0.022 \\
        & RWW & median & 0.249 ± 0.009 & 0.5 & 0.247 ± 0.018 & 0.5 & 0.250 ± 0.001 \\
        & NF + RWW & mid & 0.249 ± 0.004 & mid & \textbf{0.260 ± 0.007} & mid & \textbf{0.259 ± 0.006} \\
        
        \hline
        
        \multirow{2}{*}{GAT}
        & NF & - & \textbf{0.264 ± 0.014} & - & 0.264 ± 0.014 & - & \textbf{0.264 ± 0.014} \\
        & RWW & 0.5 & 0.255 ± 0.003 & mid & \textbf{0.298 ± 0.024} & median & 0.233 ± 0.006 \\
        & NF + RWW & mean & 0.227 ± 0.01 & mean & 0.255 ± 0.028 & median & 0.247 ± 0.015 \\
        
        \hline
        
        \multirow{2}{*}{GIN}
        & NF & - & \underline{\textbf{0.351 ± 0.09}} & - & \underline{\textbf{0.351 ± 0.09}} & - & \underline{\textbf{0.351 ± 0.09}} \\
        & RWW & 0.5 & 0.317 ± 0.053 & mean & 0.260 ± 0.003 & 0.5 & 0.272 ± 0.030 \\
        & NF + RWW & median & 0.305 ± 0.03 & 0.5 & 0.280 ± 0.028 & mid & 0.338 ± 0.051 \\
        
        \hline
        
        \multirow{2}{*}{SAGE}
        & NF & - & \textbf{0.320 ± 0.061} & - & \textbf{0.320 ± 0.061} & - & \textbf{0.320 ± 0.061} \\
        & RWW & 0.5 & 0.266 ± 0.03 & median & 0.282 ± 0.025 & mid & 0.274 ± 0.019 \\
        & NF + RWW & 0.5 & 0.295 ± 0.011 & median & 0.267 ± 0.026 & median & 0.254 ± 0.003 \\
        
    \end{tabular}
    \end{adjustbox}
    \label{tab:multiclass_macro_f1}
\end{table}

The insights above suggest that RWW-based methods improve performance in terms of accuracy. Additionally, we find that degree is an effective parameter for random weighted walks, and the median is a suitable threshold. However, we observe a drop in F1-scores, likely due to the models' difficulty in classifying graphs associated with labels that have fewer samples.

\section{Conclusion}
\label{sec:conclusion}

We propose DECODE, a density-based random weighted walk (RWW) approach that leverages local density metrics such as degree, core number, and truss number to detect coordinated campaigns in engagement networks. We prioritize density over other structural properties, as campaign graphs are consistently denser than non-campaign graphs, exhibiting higher mean degree, core number, and truss number. DECODE learns density-aware embeddings using RWWs, where node transitions are guided by local density, ensuring that neighboring nodes have similar density characteristics. These RWWs are then converted into density-aware embeddings using Skipgram.
We train an MPNN using these embeddings on the LEN dataset and observe performance improvements, surpassing the accuracy of \cite{Gopalakrishnan2025LargeEN} by 11\% and 4.5\% in binary and multiclass classification, respectively. Additionally, we outperform their F1-score for binary classification by 0.112. However, our highest macro-F1 score for campaign type classification is 0.013 lower than the best-performing baseline from Gopalakrishnan et al. This is due to the label disparity issues in the campaign classification problem.
For future work, we aim to explore alternative RWW methods, such as nearest-neighbor RWW, instead of thresholding approaches. Additionally, we plan to incorporate other structural properties, such as betweeness centrality and clustering coefficient, to further refine the RWW process.

\begin{credits}
    \subsubsection{\ackname} A. A. Gopalakrishnan, J. Hossain, and A. E. Sariyuce are supported by NSF awards OAC-2107089 and IIS-2236789, and this research used resources from the Center for Computational Research at the University at Buffalo (CCR 2025) ~\cite{CCR}.
\end{credits}

%
%
%
%


\end{document}